\magnification=\magstep1
\tolerance=500
\bigskip
\rightline{14 July, 2016}
\bigskip
\centerline{\bf A Statistical Mechanical Model for Mass Stability}
\centerline{\bf  in the} 
\centerline{\bf SHP Theory}
\bigskip
\centerline{Lawrence P. Horwitz}
\smallskip
\centerline{School of Physics, Tel Aviv University, Ramat Aviv, Israel}
\centerline{Department of Physics, Bar Ilan University, Ramat Gan, Israel}
\centerline{Department of Physics, Ariel University, Ariel, Israel}
\bigskip
\noindent {\it Abstract}: We construct a model for a particle in the framework of the theory of Stueckelberg, Horwitz and Piron (SHP) as an ensemble of events subject to the laws of covariant classical equilibrium statistical mechanics. The canonical and grand canonical ensembles are constructed without an {\it a priori} constraint on the total mass of the system. We show that the total mass of the system, corresponding the mass of this particle is determined by a chemical potential.  This model has the property that under perturbation, such as collisions in the SHP theory for which the final asymptotic mass of an elementary event is not constrained by the basic theory, the particle returns to its equilibrium mass value. A mechanism similar to the Maxwell construction for more than one equilibrium mass state may result in several possible masses in the final state. 
\bigskip
\noindent{\bf 1. Introduction}
\bigskip
\par The relativistic classical and quantum mechanics based on the theory of Stueckelberg, Horwitz and Piron [SHP] [1],[2],[3] admits a wide range of values after collision for the quantity $p_\mu p^\mu = {\bf p}^2 - E^2 = -m^2$ for each of the particles, bounded only by kinematics and the conserved value of, for example, the Hamiltonian of a system of two particles in a potential type interaction, 
$$K= {{p_1}_\mu {p_1}^\mu\over 2M_1 }+ {{p_2}_\mu {p_2}^\mu\over 2M_2 }+ V(x_1 -x_2), \eqno(1.1)$$
where $M_1$ and $M_2$ are intrinsic parameters with dimension mass for each of the particles (these quantities may be thought of as the Galilean limiting masses, as discussed in ref. [4]), and $x\equiv \{x^\mu\}$, the spacetime coordinates of the {\it events} that are considered the basic dynamical objects of the theory (we shall call these {\it particles}, although the identification of these objects with particle in the usual sense, as discussed in [5] and in Jackson's book [6] involves the construction of conserved  currents).  The equations of motion, according to  the Hamilton equations derived from this Hamiltonian, describe evolution in an an invariant parameter $\tau$[1][2][3].  Solving the equations of motion for an asymptotically free ($V\cong 0$) incoming state with well-defined incoming mass values, in the outgoing asymptotic state the particles may have different mass values. In order to maintain the identity of particles in such asymptotic states, it has been assumed in many applications, that there is a mechanism, such as relaxation to a minimum free energy, which restores the particle masses ({\it e.g.}, at the ionization point for the bound state problem [7]).  For example, an electron suffering many collisions, even though the mass changes may be small in each collision, could drift measureably, in contradiction with the observed stability of the electron mass under a wide range of conditions. Although the assumption of some relaxation process has been adequate to obtain quantitative results in agreement with experiment, the availability of an effective physical model could have important predictive consequences[8]
\par In this paper, I formulate a model for an ``elementary'' Stueckelberg particle by constructing a statistical ensemble of the type studied by Horwitz, Schieve and Piron [4],where the relativistic Gibbs ensembles for a system of events in thermal equilibrium were worked out. In order to achieve a structure for which the nonrelativistic limit provided the correct {\it a priori} nonrelativistic limit for the total mass of the system, they assumed that the value of the many-body Hamiltonian was restricted, in the microcanonical ensemble, to this value of total nonrelativistic mass. In the model I construct here, the total mass value of the ensemble is not restricted; it may vary according to the dynamical Hamiltonian laws of motion laws of motion as described above,  but in the asymptotically free state, the mass value is determined, as for the average energy, by a chemical potential analogous to a temperature (one could think of the associated relaxation as accompanied by emission or absorption of radiation).  The statistical ensemble therefore has the property that its mean spacetime value behaves under interaction as the usual Stueckelberg particle, but contains a mechanism for stability of its mass after interaction, with the possibility of transitions to other equilibrium states according to the Maxwell construction, as a phase transition induced by the interactions. The fundamental origin of such an ensemble may lie in the vacuum fluctuations of a quantum field theory, but we leave this for future study. 
\bigskip
\noindent{\bf 2. SHP Dynamics and Statistical Mechanics: The Model}
\bigskip
\par For a single particle, Stueckelberg [1]considered the possibility that the world line of a particle might turn from advancing in $t$ to a state in which the motion in $t$ is reversed, and recognized that this situation describes pair annihilation in classical mechanics. Since this motion is not single valued in $t$, he defined a parameter $\tau$ along the trajectory.  In terms of this parameter, he was able to define a canonical Hamiltonian formalism in spacetime. Horwitz and Piron [2] then generalized the structure of the theory, postulating that $\tau$ is universal (as for $t$ in Newtonian-Galilean mechanics),  in order to deal with the many-body problem (see ref. 3 for a discussion of applications and consequences of this idea). The notion of $\tau$ evolution of a many body system permits us to think of equibrium ensembles for a many body system [4]. In the following we formulate the statistical mechanics of a system of many degrees of freedom that we imagine to be associated with the structure of a {\it single Stueckelberg particle} to find equlibrium conditions for the stability of its mass.  
\par The relativistically invariant Hamiltonian for an $N$-body system of particles interacting through a Poincar\'e invariant potential is defined [2] by
$$ K = \Sigma_{i= 1}^N {{p_i}_\mu {p_i}^\mu \over 2M_i } + V(x_1.x_2, \dots, x_N) \eqno(2.1)$$

\par We imagine that this set of particles forms an ensemble which plays the dynamical role of a {\it single} 
Stueckelberg particle with dynamical coordinates $P^\mu, X^\mu$ in phase space, corresponding to the center-of-mass properties of the ensemble. We then define the microcanonical ensemble [4] as
$$\Gamma(\kappa, E) = \int d\Omega \delta (K-\kappa) \delta (\Sigma E_i -E), \eqno(2.2)$$
where $d\Omega $ is the infinitesimal volume element in the phase space of the many-body system. We do not choose {\it a priori}, however, the parameter $\kappa$ to correspond to the total Galilean mass of the system, but leave its eventual value to equilibrium conditions imposed, as for $E$, by a chemical potential ( a ``mass temperature''). 
\par The canonical ensemble is then constructed by extracting a small subensemble for which the energy and mass parameter are assumed to be additive as in the usual statistical mechanics.  This procedure distinguishes what we shall think of as the ``particle'' (the subensemble) from its ``environment'' (the ``bath''). We write the microcanonical ensemble in terms of a sum over all possible partitions of energy and mass parameter beteen the system and the bath (one usually thinks of an exchange of energy between the subsystems and the bath; in this case, we include also an exchange of mass, reflecting not the flow of heat only, but also of off-shell mass as an intrinsic degree of freedom):
$$\eqalign{\Gamma (\kappa, E)&=\int d\Omega_b d\Omega_s d\kappa_b d\kappa_s \cr
&\delta (K_b - \kappa_b) \delta(K_s - \kappa_s)\delta (E_s +E_b -E) \delta (\kappa_s +\kappa_b -\kappa) \cr
&= \int d\kappa_s d\Omega_s \Gamma_b(\kappa-\kappa_s, E-E_s) \delta (K_s -\kappa_s) \cr
&= \int d\kappa' dE' d\kappa_s d\Omega_s \Gamma_b (\kappa-\kappa', E-E')\delta (\kappa_s-\kappa')\delta(E_s-E')\cr
&= \int d\kappa' dE' \Gamma_b(\kappa-\kappa',E-E') \Gamma_s(\kappa',E'),\cr} \eqno(2.3)$$
where the subscript $b$ corresponds to the bath, and $s$ to the subsystem.\footnote{*}{ We remark that the integration
$\int d\kappa \Gamma(\kappa,E)$ is convergent and maximizing the remaining integrand over $E'$ would provide just a single temperature $T$.}
\par We then suppose that the integrand has a maximum over both variables $\kappa', E'$.  In case there is more than one local maximum in either variable, the usual Maxwell construction will yield, in addition     
to several equilibrium energy states, possibly several equilibrium mass states (as, for example, in the muon-electron system). The conditions for a maximum in two variables contain the necessary condition that there is a vanishing derivative in each of the variables. This condition is sufficient to construct the associated thermodynamics. The remaining requirmement for a maximum on the two variables is consistently automatically satisfied in the neighborhood of this maximum by the additivity property of the entropy, as we demonstrate in the following. 
\par In addition to the necessary condition of the vanishing of the derivative of the function in each of the two variables (in our case, $E'$ and $\kappa'$ ), the eigenvalues of the matrix  ${\partial^2 f \over \partial x_i\partial x_j}$,
where 
$$f(x_1,x_2)= g(x_1,x_2)h(x_1,x_2) \eqno(2.4)$$,
and
$$\eqalign{ g(E',\kappa') &= \Gamma_s(E',\kappa')\cr 
h(E',\kappa') &= \Gamma_b(E-E', \kappa-\kappa'),\cr} \eqno(2.5)$$
must be non-positive. The quadratic equation for these eigenvalues yields non-positive roots 
only if for the first term of the solution of the quadratic, 
$$ f_{11} + f_{22} \leq 0, \eqno(2.6)$$
from which one obtains
$$ (\ln g)_{11} +(\ln h)_{11} + (\ln g)_{22} +(\ln h)_{22} \leq 0. \eqno(2.7)$$
For the second, square root term, with a little manipulation, one finds for the condition that it cannot, with the positive sign, outweigh the first term in magnitude, 
$$ \eqalign{\bigl((\ln g)_{12} + (\ln h)_{12}\bigr)^2 &\leq \bigl( (\ln g)_{11} + (\ln h)_{11} \bigr)\cr &\times \bigl( (\ln g)_{22} + (\ln h)_{22} \bigr)\cr} \eqno(2.8)$$
Now, as we shall discuss in Section 3 (see $(3.3)$), the (constant in the neighborhood of the maximum) total entropy is, by additivity,
$$ S\cong S_b +S_s = \ln h + \ln g, \eqno(2.9)$$
so that 
$$ (\ln h)_{ij} = - (\ln g)_{ij}, \eqno(2.10)$$
and therefore $(2.7)$ and $(2.8)$ are identically satisfied. Therefore, in the neighborhood of the assumed maximum, the condition for the absence of a saddle point configuration is consistently satisfied.  
\par We emphasize again that our conditions are local, and that there may be more than one such maximum in the microcanonical distribution, as in the formulation of the usual statistical mechanics, with more than one mass state comprising different phases.
  \par The conditions for equilibrium can therefore be written
$$ {1 \over \Gamma_b(\kappa-\kappa',E-E')} {\partial \Gamma_b  \over \partial E}(\kappa-\kappa', E-E')|_{max}
 = {1\over \Gamma_s(\kappa',E')}{\partial \Gamma_s\over \partial E}(\kappa',E')|_{max}\equiv {1 \over T} \eqno(2.11)$$
and
$$ {1 \over \Gamma_b(\kappa-\kappa',E-E')} {\partial \Gamma_b  \over \partial \kappa}(\kappa-\kappa',E-E')|_{max}= {1\over \Gamma_s(\kappa',E')}{\partial \Gamma_s\over \partial \kappa}(\kappa',E)|_{max}
\equiv {1 \over T_\kappa}, \eqno(2.12)$$
defining a new effective ``mass temperature''.
\bigskip
\noindent{\bf 3. Canonical Thermodynamics}
\bigskip
\par Let us now define the entropies of the bath and the subsystem,
$$ \eqalign{S_b(\kappa,E) &= \ln \Gamma_b(\kappa,E)\cr
S_s(\kappa, E)&= \ln \Gamma_s(\kappa,E), \cr} \eqno(3.1)$$
 so that (we take the Boltzmann constant $k_B$, which we set equal to unity, the same for both definitions to assure additivity) at maximum, according to $(2.4)$ and $(2.5)$,\footnote{*}{Note that $K$ is actually proportional to a negative mass in our metric, so that $-T_\kappa$ is a positive number, to be identified with a ``mass temperature''.} 
$$\eqalign{   {\partial S_b \over \partial E} &= {\partial S_s \over \partial E}  = {1 \over T}\cr
     {\partial S_b \over \partial \kappa} &= {\partial S_s \over \partial \kappa}  = {1 \over T_\kappa}.\cr}\eqno(3.2)$$
The total entropy of the system is then,  independent of $\kappa',E'$ in the neighborhood of the maximum,
$$ S(\kappa,E) \cong \ln \Gamma_b(\kappa-\kappa',E-E') + \ln \Gamma_s(\kappa',E') \eqno(3.3)$$
\par In this neighborhood we  may define 
$$ \Gamma_b(\kappa-\kappa',E-E') = e^{S_b(\kappa-\kappa',E-E')}. \eqno(3.4)$$
For $\kappa'$ and $E'$ small compared to $\kappa$ and $E$,
$$ \eqalign{\Gamma_b(\kappa-\kappa',E-E') &= e^{S_b(\kappa-\kappa',E-E')}\cr
&\cong  e^{S_b(\kappa,E)-\kappa' {\partial S_b \over \partial \kappa} -E' {\partial S_b \over \partial E}}\cr
&= e^{S_b(\kappa,E)} e^{-{\kappa' \over T_\kappa}} e^{- {E' \over T}}; \cr}\eqno(3.5)$$
We then have, integrating over $\kappa'$ and $E'$ (essentially restricted to the neighborhood of the maximum) 
$$ \Gamma(\kappa,E) = \int d\kappa' dE' e^{S_b(\kappa,E)}e^{ -\kappa' {\partial S_b \over \partial \kappa}} e^{ -E' {\partial S_b \over \partial E}} \Gamma_s(\kappa', E'), \eqno(3.6)$$
Recall  that
$$\Gamma_s(\kappa',E') = \int d\Omega_s \delta (K_s -\kappa') \delta(E_s - E') \eqno(3.7)$$
and therefore we may write
$$\eqalign{\Gamma(\kappa,E) &=\int d\Omega_s d\kappa' dE' \delta(K_s-\kappa')\delta(E_s - E')\cr
&e^{S_b(\kappa,E)} e^{-{\kappa' \over T_\kappa}} e^{- {E' \over T}} \cr
&= e^{S_b(\kappa,E)}\int d\Omega_s e^{-{K_s \over T_\kappa}}e^{-{E_s \over T}}. \cr} \eqno(3.8)$$
Since $S_b(\kappa,E)$ is an overall factor, it cancels out in any computation of average values. The distribution function for the $N$-body canonical ensemble is then (we drop the subscript $s$)  
  $$ d_N(q,p) = e^{-{K \over T_\kappa}}e^{-{E \over T}} \eqno(3.9)$$
and define the partition function as
$$ Q_N (T_\kappa, T) = \int d\Omega e^{-{K \over T_\kappa}}e^{-{E \over T}}, \eqno(3.10)$$
a very different result from [4], since $K$ is considered here (as for $E$) as a dynamical variable.
\par Let us now define the Helmholtz free energy $A$ by
 $$ Q_N (T_\kappa, T) = e^{-A(T_\kappa,T) \beta}, \eqno(3.11)$$
so  that (following Huang[9]), we write 
$$ \int d\Omega e^{-\beta_\kappa K } e^{(A-E)\beta} = 1, \eqno(3.12)$$
where $\beta = 1/T$ and $\beta_\kappa = 1/T_\kappa$. Differentiating with respect to $\beta$ one obtains
$$0 = \int d\Omega e^{-\beta_\kappa K} (A-E + \beta {\partial A \over \partial\beta}) e^{(A-E)\beta}. \eqno(3.13)$$
Using $(3.10)$ for the definition of $Q_N$, we see that
$$ \eqalign{A&= <E> -\beta {\partial A \over \partial \beta}\cr
&= <E> + T{\partial A \over \partial T}. \cr} \eqno(3.14)$$
Comparing this result with what we know of thermodynamics, if we define
$$ S = - {\partial A \over \partial T}, \eqno(3.15)$$
we have 
$$ A = <E> -TS. \eqno(3.16)$$
The formula $(3.15)$ can be derived, as we shall see, from the grand canonical ensemble.
\par There  is another relation, however, that we can obtain by taking the derivative of $(3.12)$ with respect to $\beta_\kappa$; one obtains in this way
$$\eqalign{ 0 &= \int d\Omega \bigl\{ -K e^{-\beta_\kappa K}  + e^{-\beta_\kappa K} \beta {\partial A \over \beta_\kappa}\bigr\} e^{(A-E)\beta}\cr
&= -<K> + \beta {\partial A \over \partial \beta_\kappa},\cr} \eqno(3.17)$$
 so that
$$ <K>= {1\over T} {\partial A \over \partial \beta_\kappa} = -{T_\kappa^2 \over T}{\partial A \over \partial T_\kappa}, \eqno(3.18)$$
which approaches a limiting value of $-{1 \over 2} Mc^2$ in the nonerelativistic limit, as required in [4].   
\par We therefore obtain a mean value for $<K>$, the effective center-of-mass mass of the subensemble,  which is determined by $T_\kappa$ and $T$, under the canonical distribution, corresponding to an equlibrium of both heat and mass, without exchange of particles with the bath.\footnote{*}{Note that the Particle Physics Booklet,[10], in the discussion of big bang cosmology, page 236, summarizes deconfinement temperatures for several quarks and particles.}
\par We further remark that from $(3.14)$,
$$ {\partial A \over \partial \beta_\kappa} = {\partial <E> \over \partial \beta_\kappa} - \beta {\partial^2 A\over \partial \beta \partial \eta_\kappa};\eqno(3.19)$$
so that, by $(3.18)$,
$$<K> = \beta {\partial <E> \over \partial \beta_\kappa} - \beta^2 {\partial^2 A\over \partial \beta \partial \eta_\kappa} \eqno(3.19)$$
This result emphasizes the relation between the dynamical veriables $E$ and $K$,
\bigskip
\noindent{\bf 4. Fluctuations in Mass and Energy}
\bigskip
\par To compute the fluctuations in energy, we again follow the efficient method used in Huang[9], and take the derivative with respect to $\beta$ of (from $(3.12)$)
 $$ 0 = \int d\Omega e^{-\beta_\kappa K} (E-<E>) e^{(A-E)\beta}. \eqno(4.1)$$
Using, from $(3.14)$, the result that $A + {\partial A \over \partial \beta}\beta = <E>$, one obtains 
$$ < (E-<E>)^2> = - {\partial <E> \over \partial \beta} = T^2 {\partial <E> \over \partial T} \eqno(4.2)$$
As for the usual statistical mechanics, we can define at constant space volume $V^{(3)}$, implicitly assumed here,
$$ C_V = \bigl({\partial <E> \over \partial T}\bigr)_V \eqno(4.3)$$
In the same way, the mass fluctuations can be computed by considering
$$0 = \int d\Omega e^{-\beta_\kappa K} (K-<K>) e^{(A-E)\beta}. \eqno(4.4)$$ 
Differentiating with respect to $\beta_\kappa$, and using $(3.18)$, one obtains a result analogous to $(4.2)$, 
$$<(K-<K>)^2> = - {\partial <K> \over \partial \beta_\kappa}, \eqno(4.5)$$
implying that ${\partial <K> \over \partial \beta_\kappa} <0$, or, since the right side is $+{T_\kappa}^2   
 {\partial <K> \over \partial T_\kappa}$, that 
$${\partial <K> \over \partial T_\kappa} > 0, \eqno(4.6)$$ 
{\it i.e.}the mean mass (negative $K$)rises with the ``mass temperature''(negative $T_\kappa$). We note that, furthermore, by $(3.18)$,
$$  {\partial <K> \over \partial \beta_\kappa} = \beta {\partial^2 A \over {\partial \beta_\kappa}^2}.\eqno(4.7)$$
These relations provide an interesting starting point for the development of the thermodynamics of the mass of our model particle, which will be further developed elsewhere. 
\par In addition to the mean square mass and energy fluctuations, it is interesting to compute the cross correlation.  This can be achieved by differentiating $(4.4)$ with respect to $\beta$ (or $(4.1)$ with respect to $\beta_\kappa$), and yields in the same way the relation
$$ \eqalign{<(E-<E>)(K-<K>)> &= <EK> -<E><K> \cr
&= - {\partial <K> \over \partial \beta} = -{\partial <E> \over \partial \beta_\kappa},\cr } \eqno(4.8)$$
so that we have the symmetrical result 
$${\partial <K> \over \partial \beta} = {\partial <E> \over \partial \beta_\kappa}.\eqno(4.9)$$
This last result also follows from the continuity of the free energy in $\beta$ and $\beta_\kappa$ by differentiating $(3.14)$ with respect to $\beta_\kappa$ and $(3.18)$ with respect to $\beta$. 
\par We remark that the bracket type formula
$$\eqalign{\{<E>,<K>\}&= {\partial <E> \over \partial \beta}{\partial <K> \over \partial \beta_\kappa}
-{\partial <K> \over \partial \beta}{\partial <E> \over \partial \beta_\kappa}\cr
&= < (E-<E>)^2><(K-<K>)^2> \cr
&- <(E-<E>)(K-<K>)>^2 \cr} \eqno(4.10)$$
vanishes in the absence of correlation,\footnote{*}{It clearly vanishes in the nonrelativistic limit, but also in the relativistic on-shell theory, {\it i.e.}, for which $E= \sqrt{{\bf p}^2 +m^2}$.} and thus carries a measure of the correlation of fluctuations in $E$ and $K$ in the relativistic canonical ensemble.
\bigskip
\noindent{\bf 5.Grand Canonical Ensemble}
\bigskip
\par We recognize that the results above were obtained for a given number $N$ of events in the ensemble, a picture that is not very realistic in view of our interpretation that the source of the ensemble might lie in the vacuum fluctuations of a quantum field theory. In this section, we allow the number of events in the ensemble to vary (exchange with the bath) as well as its volume, and construct the grand canonical ensemble.  We write the decomposition of the full microcanonical in terms of its canonical subsets
$$ \eqalign{Q_N (V,T,T_\kappa) &=  \Sigma_{N_s=0}^N \int d\Omega_s e^{-\beta_\kappa K_s}e^{-\beta E_s}\cr
&\times Q_{N-N_s}(V-V_s, T, T_\kappa), \cr }\eqno(5.1)$$
where $s$ refers to the $N_s$ body subsystem ($T$ and $T_\kappa$ are equilibrium parameters for the whole system), and, for the bath at each $N_s$,
$$Q_{N-N_s}(V-V_s, T, T_\kappa)= \int d\Omega_b e^{-\beta_\kappa K_b} e^{-\beta E_b}, \eqno(5.2)$$
where $K_b= K-K_s$ and $E_b = E-E_s$. The normalized phase space density is then
$$ \eqalign{\rho(\Omega_s, N_s) &= {1 \over Q_N (V,T,T_\kappa)}e^{-\beta_\kappa K_s}e^{-\beta E_s}\cr
&\times Q_{N-N_s}(V-V_s, T, T_\kappa), \cr }. \eqno(5.3)$$
We can now write (for $N_s$ small compared to $N$, which can be arbitrarily large),
$$ \eqalign{Q_{N-N_s}(V-V_s, T, T_\kappa) &= e^{-\beta A(V-V_s, T, T_\kappa, K-K_s, N-N_s)}\cr
  \cong Q_N (V,T,T_\kappa)e^{\beta V_s {\partial A\over \partial V} +
{\partial A \over \partial K}K_s + {\partial A \over \partial N} N_s}\cr}. \eqno(5.4)$$
We then use the usual identifications with the pressure and chemical potential for the number
$$\eqalign{ {\partial A \over \partial V} &= -P \cr
 {\partial A \over \partial N}&= \mu\cr} \eqno(5.5)$$
and define the new mass chemical potential
$$ {\partial A \over \partial K} = -\mu_\kappa \eqno(5.6)$$
The microcanonial partition function can now be written as (with $N$ arbitrarily large)
$$\eqalign{Q_N(V,T,T_\kappa) &= \Sigma_{N_s =0}^N \int\Omega_s e^{-\beta_\kappa K_S} e^{-\beta E_s}\cr
&\times e^{-\beta V_s E - \mu_\kappa \beta K_s + \beta \mu N_s}.\cr} \eqno(5.7)$$
We note that $K_s$ appears multiplied by a linear combination of $\mu_\kappa \beta$ and $\beta_\kappa$, arising from an interplay between the equilibrium requirements of the canonical ensemble (allowing variable mass) and the structure of the grand canonical ensemble (allowing varying particle number). Let us define
$$ {\hat \mu}_\kappa = \mu_\kappa + {\beta_\kappa \over \beta} = \mu_\kappa + {T \over T_\kappa}.\eqno(5.8)$$
 We shall, furthermore, assume that $V_s$ has the common value $V$ for every subensemble. The grand partition function can then be defined as
 $${\cal Q}(V, T,T_\kappa) = e^{\beta V P} = \Sigma_{N_s0}^N z^{N_s} Q_{N_s} (T, K_s,E_s), \eqno(5.9)$$
where
$$  Q_{N_s} (T, K_s,E_s)= \int d\Omega_s \zeta^{K_s} e^{-\beta E_s} \eqno(5.10)$$
and
$$ \eqalign{z&= e^{\mu \beta},\cr
 \zeta&=e^{-{\hat \mu}_\kappa \beta}.\cr} \eqno(5.11)$$
It then follows that
$$ <N> = z {\partial \over \partial z} \ln {\cal Q} \eqno(5.12)$$
and, for the mean grand canonical mass,
$$ <K> = \zeta {\partial \over \partial \zeta} \ln {\cal Q}. \eqno(5.13)$$  
\bigskip
\noindent{\bf 6. Grand Canonical Thermodynamics}
\bigskip
\par In this section, we make a connection with entropy and thermodynamics, and show consistency with our earlier definition of entropy in the canonical ensemble,
\par We define the Helmholtz free energy for the grand canonical ensemble, modified from the usual definition by taking into account the additional mass degree of freedom, 
$$A = {<N> \over \beta} \ln z + {<K> \over \beta} \ln \zeta - { 1 \over \beta} \ln {\cal Q}\eqno(6.1)$$
so that
$${\cal Q}= e^{-\beta A} z^{<N>} \zeta^{<K>} \eqno(6.2)$$
From $(5.9)$, $(5.10)$, we then see, replacing $N_s$ by $N$ and taking the sum to infinity, that
$$ 1 = \Sigma_{N=0}^\infty z^{N-<N>} \int d\Omega \zeta^{K-<K>} e^{-\beta(E-A)}. \eqno(6.3)$$
Differentiating with respect to $\beta$ ( holding $z,\zeta,V$ fixed), one obtains
$${\partial \over \partial \beta} (\beta A)= <E> + {\partial <N> \over \partial \beta} \ln z + 
{\partial <K> \over \partial \beta} \ln \zeta. \eqno(6.4)$$
Now, by $(6.1)$,
$$\beta A = <N> \ln z + <K> \ln \zeta -\ln {\cal Q}, \eqno(6.5)$$
so that
$$ {\partial \over \partial \beta} (\beta A)={\partial <N> \over \partial \beta} \ln z + 
{\partial <K> \over \partial \beta} \ln \zeta-{\partial \over \partial \beta} \ln {\cal Q}.\eqno(6.6)$$
With $(6.4)$, we find the result for the internal energy
$$ U \equiv <E> = -{\partial \over \partial \beta} \ln {\cal Q}. \eqno(6.7)$$
Now, add and subtract $(6.5)$ (divided by $\beta$) to make the dependence on $A$ explicit, 
we have, with the definitions of $z$ and $\zeta$  in terms of the chemical potentials,
$$U= A -<N> \mu + {\hat \mu}_\kappa <K> + {1 \over \beta} \ln {\cal Q} -{\partial \over \beta} \ln {\cal Q}.\eqno(6.8)$$
Using
$$  -{\partial \over \beta} \ln {\cal Q} = + T^2 {\partial \over \partial T}\ln {\cal Q}\eqno(6.9)$$  
and using the thermodynamic relation
$$ U = A + TS, \eqno(6.10)$$
one finds
$$S = {\partial\over \partial T} (T \ln {\cal Q}) + {{\hat \mu}_\kappa \over T} <K> - {\mu \over T} <N>. \eqno(6.11)$$
With the definition $(5.8)$, we can write this as
$$S = {\partial\over \partial T} (T \ln {\cal Q} + \bigl({\mu_\kappa \over T}+ {1\over T_\kappa}\bigr) <K> - {\mu \over T} <N>. \eqno(6.11)$$
\par We can now obtain the Maxwell relations by taking the differential of  $(6.1)$ and using
$${\partial \over \partial V} \ln{\cal Q} = {P\over T} \eqno(6.12)$$
to obtain
$$ \eqalign{dA &= T\ln z d<N> +T\ln \zeta  d<E> + \bigl\{ {<N> \over T}\mu\cr
&- {<K>  \over T}{\hat \mu}_\kappa - {\partial \over \partial T} (T \ln {\cal Q}) \bigr\} dT -PdV.\cr} \eqno(6.13)$$
The terms in brackets are the negative of our definition of entropy $(6.11)$, so that we obtain the Maxwell relations
$$\eqalign{ S &=-\bigl({\partial A \over \partial T}\bigr)|_{V,<N>,<K>}  \cr
P &= -\bigl( {\partial A \over \partial V}\bigr)|_{<N>,<K>,T}\cr} \eqno(6.14)$$
as well as
$$\eqalign{ {\partial A \over \partial <N>} &= \mu\cr
{\partial A \over \partial <K>} &= - {\hat \mu}_\kappa = - (\mu_\kappa + {T \over T_\kappa})\cr} \eqno(6.15)$$
\par When the free energy reaches a critical point in $<K>$, {\it i.e.}, 
$$ {\partial A \over \partial <K>} =0,\eqno(6.16) $$
then (since $T_\kappa$ is negative, this corresponds to a positive $\mu_\kappa$).
$${T \over T_\kappa} = -\mu_\kappa. \eqno(6.17)$$
\bigskip
\noindent{\bf 7. Conclusions}
\bigskip
\par In the classical SHP theory, the fundamental dynamical object is an {\it event}, moving in spacetime according to laws of motion generated, in the case of our study here, by a Hamiltonian according to Hamiltion's equations.  The usual notion of a particle is associated with the world line of this event[3][5][6]], but for simplicity we have called this fundamental object a ``particle'' here. The mass (squared)of the particle, the negative of $p_\mu p^\mu$ (in our metric), after a collision, may asymptotically take on a wide range of values, limited by kinematics and the conserved quantity $K$, the (invariant) Hamiltonian. One sees that in experiment, the mass of a particle, such as the electron, is generally to be found very close to a definite value even after many collisions have taken place, and the question naturally srises of how this stability can be understood in the framework of the theory.
\par We have constructed, in this paper, a model for which the mass of the particle is controlled by a chemical potential, so that asymptotic variations in the mass can be restored to a given value by relaxation to satisfy the equilibrium conditions. In this model, we have taken the ``particle'' to be a statistical ensemble which has both an equilibrium energy and an equilibrium mass, controlled by the temperature and chemical potentials, thus assuring asymptotic states with the correct mass. The thermodynamic properties of such a system, involving the maximization of the integrand in the microcanonical ensemble, was worked out for such a system, where both the energy and the mass are considered to be parameters of the distribution. As we have seen in Section 6, a critical point in the free energy is made available by the interplay of the equilibrium requirements of the canonical ensemble, where the total mass of the system is considered variable (off-shell), as for the energy, and the equilibrium requirements of the grand canonical ensemble, where a chemical potential arises for the particle number.
\par In case there is more than one such maximum under the integrand of the microcanonical distribution, there may be more than one equilibrium mass state, such as in the electron muon system, with transitions between these states  corresponding to phase transitions.   
\par The study of how such a system behaves under the dynamical perturbations of collisions may be studied, in this model, through the (covariant) Boltzmann equation [11]. Such a study is left for future research. 
\bigskip
\noindent{\bf Acknowledgements}
\bigskip
\par I would like to thank Mark Davidson and M. Wagman for their critical reading of the manuscript, and Y. Strauss, J. Levitan, A. Yahalom, G. Elgressy, Y. Bachar and A. Kremer for discussions.       
\bigskip
\noindent{\bf References}
\bigskip
\frenchspacing
\item{1.} E.C.G. Stueckeleberg, Helv. Phys. Acta {\bf 14}, 588 (1961); {\bf 15} , 23 (1942).
\item{2.} L.P. Horwitz and C. Piron, Helv. Phys. Acta {\bf 46}, 316 (1973).
\item{3.} Lawrence P. Horwitz,{\it Relativistic Quantum  Mechanics},  Fundamental Theories\hfil\break in Physics {\bf 180},( Springer, Dordrecht, 2015).
\item{4.} L.P. Horwitz, W.C. Schieve and C.Piron, Ann. of Phys. {\bf 137}, 306 (1981).
\item{5.} L.P. Horwitz and Y. Lavie, Phys. Rev. {\bf D 26}819 (1982).
\item{6.} J.D. Jackson, {\it Classical Electrodynamics}, 2nd edn,(Wiley, New York 1974).
\item{7.} R.I. Arshansky and L.P. Horwitz, Jour. Math. Phys. {\bf 30}, 66, 213,380 (1989).
\item{8.} I thank Don Reed and Mark Davidson for emphasizing this question,  and to C. Piron, for first qualitatively referring to such a mechanism.
\item{9.} K. Huang, {\it Statistical Mechanics}, Wiley, New York (1967).
\item{10.} Particle Physics Booklet, July (2014), extracted from {\it Review of Particle Physics}, K.A. Olive {\it et al}, Chin. Physics C {\bf 38}, 090001 (2014).
 \item{11}. L.P. Horwitz, S. Shashoua and W.C. Schieve, Physica {\bf A 161}, 300 (1989).]

\end